\documentclass[aps,prd,twocolumn,showpacs,groupedaddress]{revtex4-1}
\usepackage{amssymb}
\usepackage{mathrsfs}
\usepackage{amsmath}
\usepackage{mathtools}
\usepackage{pstricks}
\usepackage{color}
\usepackage{graphicx}
\usepackage{array}
\usepackage{tabularx}
\usepackage{physics}
\usepackage{float}

\begin{document}


\title{\bf Nonextensive statistical field theory}



\author{P. R. S. Carvalho}
\email{prscarvalho@ufpi.edu.br}
\affiliation{\it Departamento de F\'\i sica, Universidade Federal do Piau\'\i, 64049-550, Teresina, PI, Brazil}




\begin{abstract}
We introduce a field-theoretic approach for describing the critical behavior of nonextensive systems, systems displaying global correlations among their degrees of freedom, encoded by the nonextensive parameter $q$. As some applications, we report, to our knowledge, the first analytical computation of both universal static and dynamic $q$-dependent nonextensive critical exponents for O($N$) vector models, valid for all loop orders and $|q - 1| < 1$. Then emerges the new nonextensive O($N$)$_{q}$ universality class. We employ six independent methods which furnish identical results. Particularly, the results for nonextensive 2d Ising systems, exact within the referred approximation, agree with that obtained from computer simulations, within the margin of error, as better as $q$ is closer to $1$. We argue that the present approach can be applied to all models described by extensive statistical field theory as well. The results show an interplay between global correlations and fluctuations.  
\end{abstract}


\maketitle


\par Since a nonextensive generalization of both thermodynamics \cite{book91199467} and statistical mechanics \cite{Gibbs} was proposed \cite{Tsallis1988}, many of their aspects were examined. In one of these investigations, it was shown that the nonextensive version of thermodynamics can be mapped into its extensive counterpart through a transformation of variables \cite{PhysRevLett.88.020601}. As thermodynamics permit us to explore physics just at large length scales, signatures of nonextensivity do not manifest itself at such large scales. So nonextensive effects can emerge only when all length scales are considered, whose appropriate scenario of investigation is that of statistical mechanics, particularly when we study the critical behavior of systems undergoing phase transitions \cite{ZinnJustin,Amit}. So far, some nonextensive behaviors were observed experimentally \cite{PhysRevLett.115.238301,Lutz2013,PhysRevD.91.114027}. Many other possible applications of nonextensivity were proposed, with theoretical, experimental and computational evidences \cite{Tsallis,TSALLIS1999}. From the generalization process, a parameter $q$ $\in \mathbb{R}$ arises naturally, characterizing the nonextensivity of the theory, namely superextensivity ($q < 1$) and subextensivity ($q > 1$). The parameter $q$ is interpreted as one encoding global correlations among the many degrees of freedom of the system \cite{TSALLIS2021}. The limit $q\rightarrow 1$ recovers the extensive theory. Among the applications, that involving the critical behavior of nonextensive systems undergoing continuous phase transitions have attracted great attention in recent years \cite{PhysRevE.102.012116,PhysRevE.80.051101,PhysRevE.103.042110,10.1088/1674-1137/abf8a2,PhysRevD.101.096006,e21010031,ADLI2019773}. It was discovered some qualitative agreement between the critical behavior of real systems as the so-called manganites, e.g. La$_{0.7}$Sr$_{0.3}$MnO$_{3}$ and La$_{0.60}$Y$_{0.07}$Ca$_{0.33}$MnO$_{3}$ \cite{PhysRevB.73.092401,PhysRevB.68.014404,Reis_2002,PhysRevB.66.134417}, and the nonexetnsive two-dimensional Ising model \cite{PhysRevE.102.012116,PhysRevE.80.051101} (through computational methods). Despite this great attention, among the many theoretical methods employed for studying nonextensive systems \cite{Tsallis,TSALLIS1999}, up to date, to our knowledge, there is not a field-theoretic formulation for attaining the displayed results available in the current literature. The aim of this Letter is to fill this gap by introducing such a formulation capable of furnishing results valid in the nonextensive domain. It is based on momentum-space field-theoretic renormalization group and $\epsilon$-expansion techniques, namely the nonextensive statistical field theory (NSFT). Its corresponding extensive counterpart, \emph{i. e.}, statistical field theory (SFT) \cite{PhysRevLett.28.548}, has provided very accurate results, for example to the extensive O($N$) vector model \cite{PhysRevLett.28.548}, whose nonextensive version we have to study in this Letter. When distinct systems as a fluid and a ferromagnet are characterized by the same set of critical exponents, we say that they belong to the same universality class. Universal critical exponents satisfy to some scaling relations and do not depend on the microscopic details of the system or equivalently on its nonuniversal properties as the critical temperature, form of the lattice, etc.. They depend only on universal parameters as the dimension $d$ of the system, $N$ and symmetry of some $N$-component order parameter and if the interactions among their degrees of freedom are of short- or long-range type \cite{Aharony}. In SFT, the critical indices, for O($N$) vector models, are obtained through the fluctuation properties of some self-interacting fluctuating $N$-component scalar quantum field $\phi$ \cite{ZinnJustin,Amit}. That properties are its number $N$ of components, the properties of its internal space \cite{Aharony} (the symmetry of that space for example) and the dimension $d$ of the spacetime where the field is embedded but not on the nature of the spacetime itself, as if it is curved \cite{Costa_2019} or a Lorentz-violating one \cite{Carvalho2017} for example. Aspects like short- and long-range interactions are contained in the form of the free propagators, depending on momentum through quadratic powers \cite{ZinnJustin,Amit} and other functional forms distinct from quadratic ones \cite{Aharony}, respectively. Some scaling relations depend on the effective dimension $d_{eff}$ of the system \cite{ZinnJustin,Amit}, which is given by $d_{eff} = d$ and a function $d_{eff} \equiv d_{eff}(\varsigma_{i})$ for systems with short- and long-range interactions, respectively. The parameters $\varsigma_{i}$ characterize the long-range interactions \cite{PhysRevB.72.224432,PhysRevLett.29.917}. Also, $d_{eff}$ depends on the momentum mass dimension of the momenta present in the free propagator. Its functional dependence on momenta can be quadratic ($d_{eff} = d$) or distinct from quadratic one ($d_{eff} \equiv d_{eff}(\varsigma_{i})$) and on the volume element in momentum space of the Feynman integrals \cite{PhysRevB.72.224432,PhysRevLett.29.917}. A nonextensive scalar quantum field $\phi_{q}$, for arbitrary values of $q$, was defined \cite{PhysRevLett.106.140601,1674-1137-42-5-053102}. Moreover, the resulting nonextensive quantum field theory is nonlinear. This does not permit us to apply the superposition principle to express a general solution for $\phi_{q}$ in terms of creation and annihilation operators and then to compute the corresponding free propagator. Recently, a linearized version of that theory, for $|q - 1| < 1$, was proposed \cite{PhysRevD.98.085019} for obtaining the referred propagator, where $\phi_{q}$ turns out to be its linearized extensive counterpart, namely $\phi$. Although that free propagator depends on $q$ through its effective nonextensive mass $m_{q}^{2} = qm^{2}$, the critical exponents obtained do not depend on $q$, even for all loop orders. Then, the critical indices are extensive, at least in that version. Now we have to introduce the NSFT, which yields nonextensive $q$-dependent critical exponents values. As aforementioned \cite{PhysRevD.98.085019}, the nonextensive quantum scalar field, its mass, equation of motion, Lagrangian and free propagator turn out to be their respective extensive counterparts for $|q - 1| < 1$. We expect a better result as $q$ is closer to $1$, as an effect of the linearization procedure for obtaining the free propagator.

\par The NSFT to be introduced in this Letter is obtained as an improvement of the earlier one of Ref. \cite{PhysRevD.98.085019} by now considering the nonextensive form of the statistical weight (not considered in Ref. \cite{PhysRevD.98.085019}). Then we apply the generating functional for the interacting Euclidean quantum field theory 
\begin{eqnarray}\label{huyhtrjisd}
&& Z[J] = \mathcal{N}^{-1}\left\{\exp_{q}\left[-\int d^{d}x\mathcal{L}_{int}\left(\frac{\delta}{\delta J(x)}\right)\right]\right\}^{q}\times \nonumber \\&& \int\exp\left[\frac{1}{2}\int d^{d}xd^{d}x^{\prime}J(x)G_{0}(x-x^{\prime})J(x^{\prime})\right],
\end{eqnarray}
now without any approximation on $q$, where $\exp_{q}(-x) = [1 - (1 - q)x]_{+}^{1/(1 - q)}$ is the $q$-exponential function \cite{Tsallis1988}($[y]_{+} \equiv y\theta(y)$ and $\theta(y)$ is the Heaviside step function), rather than the conventional one $e^{-x}$ \cite{ZinnJustin,Amit}. The constant $\mathcal{N}$ is determined through the normalization condition $Z[J=0] = 1$, $G_{0}^{-1}(k) = k^{2} + m^{2}$ is the free propagator of the extensive theory (it is quadratic in momentum space, so $d_{eff} = d$ and thus the nonextensive scaling relations are the same as that of the extensive theory) and $J(x)$ is some external source to generate the correlation functions and later to be vanished. In Eq. (\ref{huyhtrjisd}), we have applied the consistent form of the $q$-distribution \cite{TSALLIS1998534} (some $q$-dependent factors present in this form can be discarded because they do not alter the final result of the nonextensive critical exponents, once they represent only an overall factor in the Lagrangian of the system) and its second term is extensive, since it is related to the free propagator, which can be defined just in the extensive realm.

\par As some illustrative applications, for O($N$)-symmetric vector models, we consider $N$-component $\lambda\phi^{4}$ scalar field theories. In the first application, we compute the static nonextensive critical exponents by applying six distinct and independent renormalization group methods and $\epsilon$-expansion techniques in $d = 4 - \epsilon$ dimensions in dimensional regularization (we follow the notation of Ref. \cite{PhysRevD.98.085019}). Motivated by the just mentioned qualitative agreement between the critical behavior of manganites and nonexetnsive two-dimensional Ising model, for which a continuous transition along the ferro-paramagnetic frontier was identified for $0.5 < q < 1$ \cite{PhysRevB.68.014404}, we make $q \rightarrow 2 - q$ in the $q$-exponential function of Eq. (\ref{huyhtrjisd}) (additive duality \cite{Eur.Phys.J.Spec.Top.}), but not on its power $q$. This procedure yields a physical, normalizable probability $q$-distribution for $q < 1$, for which $S_{q}(AB) \leq S_{q}(A) + S_{q}(B)$ (this is a reasonable result, since the global correlations turns out the system more organized as its extensive counterpart). We attain the last result by applying the additive duality for $S_{q}(AB) = S_{q}(A) + S_{q}(B) + (1 - q)S_{q}(A)S_{q}(B)$ and obtain $S_{q}(AB) = S_{q}(A) + S_{q}(B) + (q - 1)S_{q}(A)S_{q}(B)$. Moreover, it implies that $1 - (q - 1)x > 0$ for $x > 0$, which is our case where $x \propto$ to the quartic interaction. Furthermore, for $|q - 1| < 1$ \cite{Tsallis1988}, $[\exp_{2 - q}(-a)]^{q} = e^{-a}\left[1 - a(a + 2)z/2 + a^{2}(3a + 4)z^{2}/24 - \cdot\cdot\cdot \right]$, where $z = q - 1$ and $a = \mathcal{H}/k_{B}T$, and thus we expect slight $q$-dependent contributions. Also, we expect nonextensive indices values being smaller than their extensive counterparts, since correlated systems are less susceptible to temperature fluctuations than uncorrelated ones. Then, the nonextensive response functions are less divergent than their extensive counterparts. For heat capacity the effect is the opposite, once for observing a unit change of temperature in a correlated system, we have to supply a larger amount of heat than for an uncorrelated one. Corresponding ideas apply to the equation of state, whose index is $\delta_{q}$. By applying the methods, we obtain that the corresponding theories are renormalizable at leading loop order (LO) for general values of $q$ and we can compute $q$-dependent critical exponents at that order. However, at next-to-leading loop order (NLO) and beyond, the theories are nonrenormalizable for any value of $q$ but they are renormalizable only for $q = 1$. Then the nonextensive indices for NLO and higher ones are that of the extensive theory. A field theory to be renormalizable at LO for general values of some parameter and nonrenormalizable for NLO and beyond for any of the values of this parameter, but otherwise to be renormalizable only for a specific value of that parameter for NLO and higher ones is a known feature in literature. This is the case, for example, of quantum field theories in curved spacetime \cite{PhysRevD.14.1965,PhysRevLett.54.2281}. The nonextensive indices values are then composed of three-level, NLO and higher ones extensive contributions (resulting from considering all length scales) and a LO $q$-dependent term (as an effect from small length scales). The latter ones represent slight contributions as expected, according to the discussion aforementioned. We have obtained identical results for the static nonextensive critical indices through the six distinct methods, showing the arbitrariness of the renormalization schemes employed and the importance of checking the results as provided by so many methods. All the methods furnish two independent identical critical indices valid for all loop orders, namely $\eta_{q}$ and $\nu_{q}$. They are given by 
\begin{eqnarray}\label{eta}
\eta_{q} = \eta - (1 - q)\frac{(N + 2)\epsilon^{2}}{2(N + 8)^{2}}, \hspace{3mm}\nu_{q} = \nu - (1 - q)\frac{(N + 2)\epsilon}{4(N + 8)}, \nonumber \\
\end{eqnarray}
where $\nu$ and $\eta$ are their corresponding all-loop extensive critical exponents values. The four remaining ones $\alpha_{q}$, $\beta_{q}$, $\gamma_{q}$ and $\delta_{q}$ are obtained through four scaling relations among them \cite{ZinnJustin,Amit}. Their Landau values, which can also be obtained through thermodynamics (they are a result of considering only large length scales), do not depend on $q$. This fact is in accord with the general result of Ref. \cite{PhysRevLett.88.020601}. We have that $\eta_{q} < \eta$, $\nu_{q} < \nu$, etc. ($\alpha_{q} > \alpha$ and $\delta_{q} > \delta$), as expected. As a check, in Table \ref{tableexponents2}, we compare the exponents numerical outcomes for some values of $q$, to $2$d ($\epsilon = 2$) nonextensive ($q \neq 1$) Ising ($N = 1$) systems, obtained from NSFT and Ref. \cite{PhysRevE.102.012116} (in this Ref., $\beta_{q}$ and $\nu_{q}$, whose $q$-dependence is slight, as expected, are available and we have obtained the remaining ones from scaling relations as well as their respective relative errors from elementary relative error theory). The extensive indices values used are that from Onsager's exact solution \cite{PhysRev.65.117}. In this sense \emph{our results for $2d$ Ising systems are exact, within the approximation of the present Letter}. The agreement is satisfactory, within the margin of error (as better as $q$ is closer to $1$, as a result of the linearization process for obtaining the free propagator), although we are leading with the expansion parameter $\epsilon = 2$. The nonextensive indices evaluated from NSFT satisfy to the scaling relations, as opposed to the ones with strong dependence on $q$ obtained from Ref. \cite{PhysRevE.80.051101} satisfying modified scaling relations with some effective dimension $d + n_{q}$ ($n_{q} = (q^{2} + q - 2)/4$). Such modified scaling relations are not expected, since the free propagator of the theory is quadratic. Also, in Ref. \cite{PhysRevE.80.051101}, $\nu_{q}$ is independent of $q$ and violates the modified $\alpha_{q}= 2 - (d + n_{q})\nu_{q}$ scaling relation. In that Ref., the exponents were obtained through computer simulations for finite-size systems. Therefore, a finite-size scaling computation from NSFT furnishes the same bulk nonextensive exponents as that obtained in this Letter, since they do not depend on the size of the system. Their $q$-dependence gives rise to the new nonextensive O($N$)$_{q}$ universality class and implies that they satisfy to the universality hypothesis, since the $q$ parameter encodes global correlations or equivalently represents effective interactions among the degrees of freedom of the system \cite{TSALLIS1999}. According to the universality hypothesis, this leads necessarily to $q$-dependent critical indices. Alternatively, we see that the nonextensive indices must depend on $q$, once the nonextensive statistical weight of Eq. (\ref{huyhtrjisd}) promotes a modification of the internal properties of the fluctuating quantum field $\phi$ thus modifying how it interacts (depending on $q$) with itself. This implies a relation between fluctuations and global correlations. This mechanism occurs in the internal space of $\phi$ and not in the spacetime where it is embedded. This leads, necessarily, to the emergence of $q$-dependent exponents.
\begin{table}[H]
\caption{Results for the static nonextensive critical exponents for some values of $q$, to $2$d ($\epsilon = 2$) nonextensive ($q \neq 1$) Ising ($N = 1$) systems, obtained from NSFT and Ref. \cite{PhysRevE.102.012116}.}
\begin{tabular}{ p{1.5cm}p{2.5cm}p{2.0cm}p{2.0cm}  }
 \hline
 \hline
 $q$ & $\alpha_{q}$ & $\beta_{q}$ & $\gamma_{q}$   \\
 \hline
 1 Exact  &   0.000 &  0.125  &  1.750          \\
 0.9  &  0.034 &  0.121 & 1.724    \\
 0.9\cite{PhysRevE.102.012116}  &  0.020$\pm$0.014 &  0.120$\pm$0.000 & 1.740$\pm$0.014    \\
 0.8  &  0.066 &  0.117 & 1.699    \\
 0.8\cite{PhysRevE.102.012116}  &  0.054$\pm$0.012 &  0.119$\pm$0.000 & 1.701$\pm$0.012    \\
 0.7  &  0.100 &  0.114 & 1.673    \\
 0.7\cite{PhysRevE.102.012116}  &  0.082$\pm$0.014 &  0.116$\pm$0.000 & 1.686$\pm$0.014    \\
 0.6  &  0.134  & 0.110 &  1.647   \\
 0.6\cite{PhysRevE.102.012116}  &  0.090$\pm$0.016 &  0.117$\pm$0.000 & 1.676$\pm$0.016    \\
 \hline
 \hline
 $q$  & $\delta_{q}$  & $\eta_{q}$  & $\nu_{q}$ \\
 \hline
 1 Exact  & 15.000  &   0.250   &  1.000  \\
 0.9  & 15.260  &  0.246 & 0.983    \\
 0.9\cite{PhysRevE.102.012116}  &  15.529$\pm$0.008 &  0.242$\pm$0.002 & 0.990$\pm$0.007    \\
 0.8  & 15.461  &  0.243 & 0.967    \\
 0.8\cite{PhysRevE.102.012116}  & 15.327$\pm$0.008  &  0.245$\pm$0.002 & 0.973$\pm$0.006    \\
 0.7  &  15.736 &  0.239 & 0.950    \\
 0.7\cite{PhysRevE.102.012116}  & 15.529$\pm$0.008  &  0.242$\pm$0.002 & 0.959$\pm$0.007    \\
 0.6  &  16.021  & 0.235 &  0.933   \\
 0.6\cite{PhysRevE.102.012116}  & 15.327$\pm$0.008  &  0.245$\pm$0.002 & 0.955$\pm$0.008    \\
 \hline
 \hline
\end{tabular}
\label{tableexponents2}
\end{table}

\par As another application, we compute the nonextensive dynamic critical exponent $z_{q}$. Its extensive counterpart value is that evaluated up to five-loop order in Ref. \cite{Adzhemyan.five}. We compare $z_{q}$ evaluated from NSFT with the respective numerical values obtained from computer simulations for $2$d ($\epsilon = 2$) nonextensive ($q \neq 1$) Ising ($N = 1$) systems in Ref. \cite{PhysRevE.102.012116} for some values of $q$. Following similar steps as for the static case, we obtain \cite{ZinnJustin} 
\begin{eqnarray}\label{z}
z_{q} = z - (1 - q)\frac{[6\ln(4/3) - 1](N + 2)}{2(N + 8)^{2}}  \epsilon^{2}.
\end{eqnarray}
In Table \ref{tab:tablez25}, once again, we observe that the agreement is satisfactory, although for an $\epsilon$-expansion that takes into account the $\epsilon = 2$ value as the expansion parameter. 
\begin{table}[H]
\caption{Results for the dynamic nonextensive critical exponent $z_{q}$, to $2$d ($\epsilon = 2$) nonextensive ($q \neq 1$) Ising ($N = 1$) systems, obtained from NSFT and Ref. \cite{PhysRevE.102.012116}.}
\begin{tabular}{ p{4.2cm}p{4.1cm} }
 \hline
 \hline
 $q$ & $z_{q}$  \\
 \hline
 1 \hspace{1.4mm}\cite{Adzhemyan.five}  &   2.140      \\
 1 \hspace{1.4mm}\cite{PhysRevE.102.012116}  &    2.240$\pm$0.007  \\
 0.9  & 2.135    \\
 0.9\cite{PhysRevE.102.012116} & 2.151$\pm$0.006    \\
 0.8  & 2.129  \\
 0.8\cite{PhysRevE.102.012116} & 2.153$\pm$0.005  \\
 0.7  & 2.124  \\
 0.7\cite{PhysRevE.102.012116} & 2.141$\pm$0.006  \\
 0.6  & 2.118  \\
 0.6\cite{PhysRevE.102.012116} & 2.151$\pm$0.007  \\
 \hline
 \hline
\end{tabular}
\label{tab:tablez25}
\end{table}

\par Now we make some predictions on the values of both static and dynamic nonextensive critical exponents for the referred systems through Tables \ref{tableexponents2N0} - \ref{tab:tablez35N3}:

\begin{table}[H]
\caption{Results for the static nonextensive critical exponents for some values of $q$, to $2$d ($\epsilon = 2$) nonextensive ($q \neq 1$) Self-avoiding random walk ($N = 0$) systems, obtained from NSFT.}
\begin{tabular}{ p{1.5cm}p{2.5cm}p{2.0cm}p{2.0cm}  }
 \hline
 \hline
 $q$ & $\alpha_{q}$ & $\beta_{q}$ & $\gamma_{q}$   \\
 \hline
 1Exact\cite{ZinnJustin}  &   0.500 &  0.078  &  1.344          \\
 0.9  &  0.524 &  0.076 & 1.325    \\
 0.8  &  0.550 &  0.073 & 1.304    \\
 0.7  &  0.574 &  0.071 & 1.284    \\
 0.6  &  0.600  & 0.067 &  1.263   \\
 \hline
 \hline
 $q$  & $\delta_{q}$  & $\eta_{q}$  & $\nu_{q}$ \\
 \hline
 1Exact\cite{ZinnJustin}  & 18.231  &   0.208   &  0.750  \\
 0.9  & 18.512  &  0.205 & 0.738    \\
 0.8  & 18.802  &  0.202 & 0.725    \\
 0.7  &  19.101 &  0.199 & 0.713    \\
 0.6  &  19.408  & 0.196 &  0.700   \\
 \hline
 \hline
\end{tabular}
\label{tableexponents2N0}
\end{table}

\begin{table}[H]
\caption{Results for the static nonextensive critical exponents for some values of $q$, to $3$d ($\epsilon = 1$) nonextensive ($q \neq 1$) Self-avoiding random walk ($N = 0$) systems, obtained from NSFT.}
\begin{tabular}{ p{1.5cm}p{2.2cm}p{2.2cm}p{2.2cm} }
 \hline
 \hline
 $q$ & $\alpha_{q}$  & $\beta_{q}$  & $\gamma_{q}$   \\
 \hline
 1 ~\cite{ZinnJustin}   &   0.235$\pm$0.003   &  0.3024$\pm$0.0008   &  1.1596$\pm$0.0020 \\
 0.9  & 0.254$\pm$0.003  &  0.2988$\pm$0.0013 & 1.1484$\pm$0.0036    \\
 0.8   & 0.273$\pm$0.003    &  0.2951$\pm$0.0013   &  1.1368$\pm$0.0036  \\
 0.7  &  0.292$\pm$0.003 &  0.2915$\pm$0.0013 & 1.1255$\pm$0.0036    \\
 0.6   & 0.310$\pm$0.003    &  0.2879$\pm$0.0013   &  1.1139$\pm$0.0036  \\
 \hline
 \hline
 $q$ & $\delta_{q}$  & $\eta_{q}$  & $\nu_{q}$   \\
 \hline
 1 ~\cite{ZinnJustin}   &   4.8343$\pm$0.0142   &  0.0284$\pm$0.0025   &  0.5882$\pm$0.0011 \\
 0.9  & 4.8434$\pm$0.0142  &  0.0268$\pm$0.0025 & 0.5820$\pm$0.0011    \\
 0.8   &  4.8519$\pm$0.0142   &  0.0253$\pm$0.0025   &  0.5757$\pm$0.0011  \\
 0.7  &  4.8611$\pm$0.0143 &  0.0237$\pm$0.0025 & 0.5695$\pm$0.0011    \\
 0.6   &  4.8697$\pm$0.0143   &  0.0222$\pm$0.0025   &  0.5632$\pm$0.0011  \\
 \hline
 \hline
\end{tabular}
\label{N03}
\end{table}

\begin{table}[H]
\caption{Results for the static nonextensive critical exponents for some values of $q$, to $3$d ($\epsilon = 1$) nonextensive ($q \neq 1$) Ising ($N = 1$) systems, obtained from NSFT.}
\begin{tabular}{ p{1.4cm}p{2.2cm}p{2.2cm}p{2.2cm} }
 \hline
 \hline
 $q$ & $\alpha_{q}$  & $\beta_{q}$  & $\gamma_{q}$   \\
 \hline
 1 ~\cite{ZinnJustin}   &   0.109$\pm$0.004   &  0.3258$\pm$0.0014   &  1.2396$\pm$0.0013 \\
 0.9  & 0.134$\pm$0.002  &  0.3209$\pm$0.0015 & 1.2245$\pm$0.0042    \\
 0.8   & 0.159$\pm$0.002    &  0.3160$\pm$0.0015   &  1.2091$\pm$0.0042  \\
 0.7  &  0.184$\pm$0.002 &  0.3111$\pm$0.0015 & 1.1939$\pm$0.0042    \\
 0.6   & 0.209$\pm$0.002    &  0.3063$\pm$0.0015   &  1.1783$\pm$0.0042  \\
 \hline
 \hline
 $q$ &  $\delta_{q}$ & $\eta_{q}$  & $\nu_{q}$   \\
 \hline
 1 ~\cite{ZinnJustin}   &   4.8055$\pm$0.0140   &  0.0335$\pm$0.0025   &  0.6304$\pm$0.0013 \\
 0.9  &  4.8162$\pm$0.0141  &  0.0316$\pm$0.0025 & 0.6221$\pm$0.0013    \\
 0.8   &  4.8264$\pm$0.0141   &  0.0298$\pm$0.0025   &  0.6137$\pm$0.0013  \\
 0.7  &  4.8371$\pm$0.0141  &  0.0279$\pm$0.0025 & 0.6054$\pm$0.0013    \\
 0.6   &  4.8474$\pm$0.0142   &  0.0261$\pm$0.0025   &  0.5971$\pm$0.0013  \\
 \hline
 \hline
\end{tabular}
\label{N13}
\end{table}

\begin{table}[H]
\caption{Results for the static nonextensive critical exponents for some values of $q$, to $3$d ($\epsilon = 1$) nonextensive ($q \neq 1$) XY ($N = 2$) systems, obtained from NSFT.}
\begin{tabular}{ p{1.4cm}p{2.2cm}p{2.2cm}p{2.2cm} }
 \hline
 \hline
 $q$ & $\alpha_{q}$  & $\beta_{q}$  & $\gamma_{q}$   \\
 \hline
 1 ~\cite{ZinnJustin}   &   -0.011$\pm$0.004   &  0.3470$\pm$0.0016   &  1.3169$\pm$0.0020 \\
 0.9  & 0.019$\pm$0.004  &  0.3412$\pm$0.0016 & 1.2985$\pm$0.0046    \\
 0.8   & 0.049$\pm$0.004    &  0.3354$\pm$0.0016   &  1.2801$\pm$0.0046  \\
 0.7  &  0.079$\pm$0.004 &  0.3296$\pm$0.0016 & 1.26184$\pm$0.0046    \\
 0.6   & 0.109$\pm$0.004    &  0.3238$\pm$0.0016   &  1.2433$\pm$0.0046  \\
 \hline
 \hline
 $q$ &  $\delta_{q}$ & $\eta_{q}$  & $\nu_{q}$   \\
 \hline
 1 ~\cite{ZinnJustin}   &   4.7949$\pm$0.0140   &  0.0354$\pm$0.0025   &  0.6703$\pm$0.0015 \\
 0.9  & 4.8061$\pm$0.0140  &  0.0334$\pm$0.0025 & 0.6603$\pm$0.0015    \\
 0.8  &  4.8173$\pm$0.0141   &  0.0314$\pm$0.0025   &  0.6503$\pm$0.0015  \\
 0.7  &  4.8286$\pm$0.0141 &  0.0294$\pm$0.0025 & 0.6403$\pm$0.0015    \\
 0.6  &   4.8400$\pm$0.0141  &  0.0274$\pm$0.0025   &  0.6303$\pm$0.0015  \\
 \hline
 \hline
\end{tabular}
\label{N23}
\end{table}

\begin{table}[H]
\caption{Results for the static nonextensive critical exponents for some values of $q$, to $3$d ($\epsilon = 1$) nonextensive ($q \neq 1$) Heisenberg ($N = 3$) systems, obtained from NSFT.}
\begin{tabular}{ p{1.4cm}p{2.2cm}p{2.2cm}p{2.2cm} }
 \hline
 \hline
 $q$ & $\alpha_{q}$  & $\beta_{q}$  & $\gamma_{q}$   \\
 \hline
 1 ~\cite{ZinnJustin}   &   -0.122$\pm$0.010   &  0.3662$\pm$0.0025   &  1.3895$\pm$0.0050 \\
 0.9  & -0.088$\pm$0.008  &  0.3607$\pm$0.0027 & 1.3686$\pm$0.0086    \\
 0.8  & -0.054$\pm$0.008    &  0.3596$\pm$0.0027   &  1.3477$\pm$0.0086  \\
 0.7  &  -0.020$\pm$0.008 &  0.3465$\pm$0.0027 & 1.3267$\pm$0.0086    \\
 0.6  & -0.015$\pm$0.008    &  0.3399$\pm$0.0027   &  1.3056$\pm$0.0086  \\
 \hline
 \hline
 $q$ & $\delta_{q}$  & $\eta_{q}$  & $\nu_{q}$   \\
 \hline
 1 ~\cite{ZinnJustin}   &   4.7943$\pm$0.0140   &  0.0355$\pm$0.0025   &  0.7073$\pm$0.0035 \\
 0.9  & 4.8061$\pm$0.0140  &  0.0334$\pm$0.0025 & 0.6959$\pm$0.0035    \\
 0.8  &  4.8173$\pm$0.0141   &  0.0314$\pm$0.0025   &  0.6846$\pm$0.0035  \\
 0.7  &  4.8292$\pm$0.0141  &  0.0293$\pm$0.0025 & 0.6732$\pm$0.0035    \\
 0.6  &  4.8411$\pm$0.0141   &  0.0272$\pm$0.0025   &  0.6618$\pm$0.0035  \\
 \hline
 \hline
\end{tabular}
\label{N33}
\end{table}

\begin{table}[H]
\caption{Results for the dynamic nonextensive critical exponent $z_{q}$ for some values of $q$, to $3$d ($\epsilon = 1$) nonextensive ($q \neq 1$) Ising ($N = 1$) systems, obtained from NSFT.}
\begin{tabular}{ p{4.2cm}p{4.1cm} }
 \hline
 \hline
 $q$ & $z_{q}$  \\
 \hline
 1 \hspace{1.4mm}\cite{Adzhemyan.five}  &   2.0235      \\
 0.9  & 2.0222    \\
 0.8  & 2.0208  \\
 0.7  & 2.0195  \\
 0.6  & 2.0181  \\
 \hline
 \hline
\end{tabular}
\label{tab:tablez35N1}
\end{table}

\begin{table}[H]
\caption{Results for the dynamic nonextensive critical exponent $z_{q}$ for some values of $q$, to $3$d ($\epsilon = 1$) nonextensive ($q \neq 1$) Heisenberg ($N = 3$) systems, obtained from NSFT.}
\begin{tabular}{ p{4.2cm}p{4.1cm} }
 \hline
 \hline
 $q$ & $z_{q}$  \\
 \hline
 1 \hspace{1.4mm}\cite{PhysRevE.100.062117}  &   2.0330      \\
 0.9  & 2.0315    \\
 0.8  & 2.0300  \\
 0.7  & 2.0285  \\
 0.6  & 2.0270  \\
 \hline
 \hline
\end{tabular}
\label{tab:tablez35N3}
\end{table}

\par For the $3d$ systems, both static and dynamic nonextensive critical exponents depend slightly on $q$, as in the $2d$ situation.

\par In summary, we have introduced a field-theoretic approach for describing the critical behavior of nonextensive systems. We have computed analytically both universal static and dynamic nonextensive critical exponents for O($N$) models, valid for all loop orders and $|q - 1| < 1$. Then, a new universality class emerged, namely the nonextensive O($N$)$_{q}$ one. Their values are composed of both three-level, NLO and higher ones extensive contributions (resulting from considering all length scales) and a LO $q$-dependent term (as an effect from small length scales), where the latter one is slight. Both $2d$ nonextensive static and dynamic critical indices are in agreement with those obtained from computer simulation for $2$d nonextensive Ising model \cite{PhysRevE.102.012116} (Tables \ref{tableexponents2} and \ref{tab:tablez25}), within the margin of error, as better as $q$ is clorer to $1$, as a consequence of the linearization process for defining the free propagator. \emph{Our results for $2d$ Ising systems are exact, within the approximation of the present Letter}. We have showed that the nonextensive critical indices are universal, satisfy to the scaling relations and reduce to their extensive values when $q\rightarrow 1$. Furthermore, we have predicted results to the values of both static and dynamic nonextensive critical exponents for some other systems through Tables \ref{tableexponents2N0} - \ref{tab:tablez35N3}. These outcomes can be compared to their counterparts obtained though computer simulations as well as from experimental measurements \cite{PhysRevB.68.014404,Reis_2002,PhysRevB.66.134417} in a near future. The NSFT opens a new route for studying critical properties of nonextensive systems undergoing continuous phase transitions. Besides, it represents a new methodology and fills the gap in the literature for such a method like its extensive counterpart, so successful for describing extensive systems \cite{PhysRevLett.28.548}. Then, it can be applied for many other extensive models as well \cite{ZinnJustin}. 

\par \textit{Acknowledgments}---PRSC would like to thank the Brazilian funding agencies CAPES and CNPq (grants: Universal-431727/2018 and Produtividade 307982/2019-0) for financial support. 

\bibliography{apstemplate}

\end{document}